\documentstyle[12pt]{article}
\def\PRL #1 #2 #3{{\sl Phys. Rev. Lett.} {\bf#1} (#2) #3}
\def\NPB #1 #2 #3{{\sl Nucl. Phys.} {\bf B#1} (#2) #3}
\def\NPBFS #1 #2 #3 #4{{\sl Nucl. Phys.} {\bf B#2} [FS#1] (#3) #4}
\def\CMP #1 #2 #3{{\sl Commun. Math. Phys.} {\bf #1} (#2) #3}
\def\PRD #1 #2 #3{{\sl Phys. Rev.} {\bf D#1} (#2) #3}
\def\PLA #1 #2 #3{{\sl Phys. Lett.} {\bf #1A} (#2) #3}
\def\PLB #1 #2 #3{{\sl Phys. Lett.} {\bf #1B} (#2) #3}
\def\JMP #1 #2 #3{{\sl J. Math. Phys.} {\bf #1} (#2) #3}
\def\PTP #1 #2 #3{{\sl Prog. Theor. Phys.} {\bf #1} (#2) #3}
\def\SPTP #1 #2 #3{{\sl Suppl. Prog. Theor. Phys.} {\bf #1} (#2) #3}
\def\AoP #1 #2 #3{{\sl Ann. of Phys.} {\bf #1} (#2) #3}
\def\PNAS #1 #2 #3{{\sl Proc. Natl. Acad. Sci. USA} {\bf #1} (#2) #3}
\def\RMP #1 #2 #3{{\sl Rev. Mod. Phys.} {\bf #1} (#2) #3}
\def\PR #1 #2 #3{{\sl Phys. Reports} {\bf #1} (#2) #3}
\def\AoM #1 #2 #3{{\sl Ann. of Math.} {\bf #1} (#2) #3}
\def\UMN #1 #2 #3{{\sl Usp. Mat. Nauk} {\bf #1} (#2) #3}
\def\FAP #1 #2 #3{{\sl Funkt. Anal. Prilozheniya} {\bf #1} (#2) #3}
\def\FAaIA #1 #2 #3{{\sl Functional Analysis and Its Application} {\bf
#1} (#2) #3}
\def\BAMS #1 #2 #3{{\sl Bull. Am. Math. Soc.} {\bf #1} (#2)
#3} \def\TAMS #1 #2 #3{{\sl Trans. Am. Math. Soc.} {\bf #1} (#2) #3}
\def\InvM #1 #2 #3{{\sl Invent. Math.} {\bf #1} (#2) #3}
\def\LMP #1 #2 #3{{\sl Letters in Math. Phys.} {\bf #1} (#2) #3}
\def\IJMPA #1 #2 #3{{\sl Int. J. Mod. Phys.} {\bf A#1} (#2) #3}
\def\AdM #1 #2 #3{{\sl Advances in Math.} {\bf #1} (#2) #3}
\def\RMaP #1 #2 #3{{\sl Reports on Math. Phys.} {\bf #1} (#2) #3}
\def\IJM #1 #2 #3{{\sl Ill. J. Math.} {\bf #1} (#2) #3}
\def\APP #1 #2 #3{{\sl Acta Phys. Polon.} {\bf #1} (#2) #3}
\def\TMP #1 #2 #3{{\sl Theor. Mat. Phys.} {\bf #1} (#2) #3}
\def\JPA #1 #2 #3{{\sl J. Physics} {\bf A#1} (#2) #3}
\def\JSM #1 #2 #3{{\sl J. Soviet Math.} {\bf #1} (#2) #3}
\def\MPLA #1 #2 #3{{\sl Mod. Phys. Lett.} {\bf A#1} (#2) #3}
\def\JETP #1 #2 #3{{\sl Sov. Phys. JETP} {\bf #1} (#2) #3}
\def\JETPL #1 #2 #3{{\sl  Sov. Phys. JETP Lett.} {\bf #1} (#2) #3}
\def\PHSA #1 #2 #3{{\sl Physica} {\bf A#1} (#2) #3}
\def\CQG #1 #2 #3{{\sl Class. Quantum Grav.} {\bf #1} (#2) #3}
\def\SJNP #1 #2 #3{{\sl Sov. J. Nucl. Phys. (Yadern.Fiz.)} {\bf #1} (#2) #3}

\newcommand{\nn}{\nonumber\\}\newcommand{\p}[1]{(\ref{#1})}

\def\a{\alpha}\def\b{\beta}\def\d{\delta}\def\e{\epsilon}
\def\k{\kappa}\def\l{\lambda}
\def\Th{\Theta}\def\th{\theta}\def\G{\Gamma}

\begin{document}
\begin{center}
{\large \bf GENERALIZED ACTION\\ PRINCIPLE FOR SUPERSTRINGS\\ AND
SUPERMEMBRANES}
\end{center}
\bigskip
\centerline
{\large \bf Dmitrij V. Volkov }
\begin{center}
NSC Kharkov Institute of Physics and Technology\\
Kharkov, Ukraine\\
e -- mail: kfti@rocket.kharkov.ua
\end{center}

\bigskip
{\small Two approaches concerning the connection of the fermionic kappa --
symmetry with the superstring world -- sheet superdiffeomorphism
transformations are discussed. The first approach is based on the twistor
-- like formulation of the superstring action and the second one on a
reformulation of the superstring and super~-~p~-~brane actions according
to the Generalized Action Principle.}

\vspace{1truecm}
\begin{enumerate}
\item
Introduction
\item
Twistor reformulation of superparticle and \\
superstring action
\item
Rheonomic approach to superstrings and\\ supermembranes
\item
Generalized action principle for heterotic superstring
\item
Resum\'e
\end{enumerate}

\section{Introduction.}

Superstrings (and supermembranes) are extended objects not only in the
bosonic directions but also in the fermionic ones. The description of the
bosonic directions of these objects can be done by the classical
geometrical methods while, on the other hand, the description of the
fermionic directions is up to now not satisfactory.

The historically first version of superstring theory proposed by Ramond,
Neveu and Schwarz (RNS) \cite{rns} is very complicated in exposing
the target space supersymmetry. On the other hand, the
celebrated Green--Schwarz (GS) formulation \cite{gs}, being manifestly
invariant in respect to the target space supersymmetry is not so
in respect to the supersymmetry on the world
sheet, the $\kappa$--symmetry being substituted for the latter.

Twistor--like reformulation of superparticle and superstring actions has
explained the origin and the geometrical meaning of the
$~\kappa$--symmetry as the world sheet superdiffeomorphism. But at the
same time the twistor--like approach to superstrings has its foundation on
the geometrical principles which have little in common with whose used in
the case of bosonic strings.

In this lecture after discussing the motivation, main ideas and
shortcomings of the twistor~--~like approach we will consider its
reformulation on the basis of the Generalized Action Principle (GAP) which
was previously applied  to the supergravity theories. It will be shown
that GAP very conveniently formalizes the geometric procedure of
minimal embedding for supersymmetric extended objects retaining at the
same time all equations and constraints of the standard twistor like
approach.

\section{Twistor reformulation of superparticle and superstring action }

If a theory has underlying local symmetries the latter exhibit themselves
in the appearance of constraints. Investigation and reformulation of
constraints may give new insight in the symmetry properties of the
considered theory and  help to find its more adequate formulation.

We begin with consideration of the simplest case of the $~s=0~$ massless
relativistic particle. Considering the twistor form
of the constraints we will get the twistor formulation of the relativistic
particle action. Then we will generalize our consideration to the cases of
the spinning particle and superparticle, which are the limiting cases of
the RNS and GS superstrings, and will get the twistor formulation for
these cases.

And lastly we will go to superstrings and do some general conclusions
which will help to understand the main part of the lecture where the
Generalized Action Principle is introduced and examplified by its
application to the heterotic string.

\subsection{Twistor reformulation of the massless $s=0$ relativistic
particle action.}

For the massless relativistic particle with spin $s=0$, constraints follow
immediately from the particle action
\begin{equation}\label{1}
\left.
\begin{array}{l}
{\cal{L}}=\frac 12 e^{-1}\dot{x}^m\dot{x}_m \\
{\cal{L}}=p_m\dot{x}^m-\frac 12 ep^2
\end{array}
\right\}
\Rightarrow ~~ \stackrel{\mbox{\rm {
\underline{constraints} }}}
{}~~~~~~~
\cases { \dot{x}^2=0 \cr
p^2=0  \cr }
\end{equation}

For $~D=3,4,6$ and $10~$  the constraints \p{1} are satisfied if
$~\dot{x}^m ~(or~~ p_m)~$ are represented in the twistor form
\begin{equation}\label{2}
\dot{x}^m=\l\G^m\l   \qquad  p_m=-\a\l\G_m\l
\end{equation}

Using \p{2}, the massless particle action can also be reformulated as
$${\cal{L}}=p_m(\dot{x}^m-\l\G^m\l) \qquad or  \qquad
{\cal{L}}=(\l\G^m\l)\dot{x}^m$$

\subsection{Spinning particle.}

The spinning particle action has the well known form \cite{veccia}
\begin{equation}\label{3}
 {\cal{L}}=p_m\dot{x}^m-\frac 12 ep^2-\frac 12 \psi^m\dot{\psi}_m-
i\xi{\psi^m}p_m
\end{equation}
which gives the constraints
\begin{equation}\label{4}
p^2=0  \qquad ~~~~~   {\psi^m}p_m=0
\end{equation}

The first of constraints \p{4} is the same as \p{2}; the second constaint
 is its superpartner.

Let us try for the both constraints the twistor representation
\begin{equation}\label{5}
p_{\a\b}=\l_\a\l_\b  \qquad
\psi^{\a\b}=\l^\a\th^\b+\th^\a\l^\b
\end{equation}
where for simplicity we consider D=3 case.

After substitution \p{5} into \p{3} it can be easily seen that on the
mass -- shell the action \p{3} becomes
\begin{equation}\label{6}
 {\cal{L}}=\l^\a\l^\b\{\dot{x}_{\a\b}-\frac i2 (\th_\a\dot{\th}_\b+
\th_\b\dot{\th}_\a)\}
\end{equation}
with the explicit target space supersymmetry and is equivalent to the
Brink -- Schwarz \cite{bs} superparticle action.

As it follows from \p{5} $\th^{\a}$ and $\l^{\a}$ are the superpartners.
This point is the most important for further generalization and
constitutes the base of twistor -- like formulations (including the
superspace ones) of superparticles, superstrings and supermembranes.

\subsection{Supersymmetrization of the twistor particle action.}

We will show now that the superspace formulation of the action \p{6} can
be get by an immediate supersymmetrization of the
action
\begin{equation}\label{7}
{\cal{L}}=p_m(\dot{x}^m-\l\G^m\l)
\end{equation}

Using superfields
$$ P_m=p_m+i\eta\rho_m ; ~~~~ X^m=x^m+i\eta\psi^m  ~~~~~
\Th_\a=\th_\a+\eta\l_\a$$
we can supersymmetrize {\it on the world -- line} each of two terms
of the action, which procedure gives the action also as the sum of two
terms
\begin{equation}\label{8}
 S=\int\,d\tau\,d\eta\,P_m({\cal{D}}X^m-i\Th\G^m{\cal{D}}\Th)
\end{equation}
$${\cal{D}}={\partial\over{\partial}\eta}+i\eta{\partial\over{\partial}
\tau}$$

It is remarkable that the action \p{8} is {\it target space} invariant as
well as local super world~--~sheet invariant.

It can be easily shown that \p{8} is equivalent to the Brink -- Schwarz
action for D=3 superparticle.

Let us briefly consider some points which follow from \p{8}
\begin{enumerate}
\item[a)]
Since the structure of constraints is essential for understanding the main
ideas of twistor~--~like approach let us consider how it is got from the
action \p{8}.

By variation of $P_{m}$ we get the equation
\begin{equation}\label{9}
 {\cal{D}}X^m-i\Th\G^m{\cal{D}}\Th=0
\end{equation}
which usually called as the "geometrodynamical" condition.

Its integrability condition
\begin{equation}\label{10}
 {\partial_\tau}X^m-i\Th\G^m{\partial_\tau}\Th=i{\cal{D}}\Th\G^m
{\cal{D}}\Th
\end{equation}
gives the constraint
\begin{equation}\label{11}
{({\partial_\tau}X^m-i\Th\G^m{\partial_\tau}\Th)}^2=0
\end{equation}

We see that the consraint \p{11} is a superfield generalization of \p{2}.
\item[b)]
The action admits n=D-2 supersymmetrization for D=3,4,6 and 10 \cite{gs92}
$$ S=\int\,d\tau\,d^n\eta\,P_m^q({{\cal{D}}_q}X^m-i\Th\G^m{{\cal{D}}_
q}\Th)$$
$${\cal{D}}_q={\partial\over{\partial}\eta^q}+i\eta^q{\partial\over
{\partial}\tau}$$
\item[c)]
It can be shown that the Lagrange multipliers in \p{8} give no propagating
degrees of freedom
\item[d)]
In component formulation superfield $\Th$ is
$\Th^{\a}=\th^{\a}+\eta^{a}\l^{\a}_{a}+\dots$ so that twistor variables
$\l$'s are the superpartners of $\th$'s
\item[e)] On the mass shell the superdiffeomorphism of \p{8}
reproduces the $\k$ -- symmetry.
\end{enumerate}

\subsection{$ \k$ -- symmetry.}

Let us consider the last point in more details.

The local fermionic symmetry, known as the $\k$ -- symmetry was introduced
in the papers \cite{k} for the superparticle and in \cite{gs} for
the superstring.

Introduction of the $\k$ -- symmetry has solved the problem of redundant
fermionic degrees of freedom, but at the same time it has contained a
number of undesirable features: obscure geometrical meaning, the ensuing
constraints are infinitely reducible, the on mass shell formulation is
only available. All these points is a handicap to covariant quantization
of superstring theories.

To get the idea how the $\k$ -- symmetry is related to the worldsheet
superdiffeomorphism we consider the $\k$ -- transformations
$$ \d\th=i(\G{p})\k \qquad \d{x^m}=i\th\G^m\d\th$$
For the simplest case $D=3$ superparticle
$$ \d\th_\a=p_{\a\b}\k^\b$$
Recalling that $ p_{\a\b}=\l_\a\l_\b$
$$ \d\th_\a=\l_\a\l_\b\k^\b=\l_\a(\l_\b\k^\b)=\l_\a\a(\k)$$
we get that $ \a(\k)$ is the odd parameter of superdiffeomorphism
transformations.

In general for D=3,4,6,10, if $ p^m=\l\G^m\l$, then one of $ \l$'s
in the $ \k$ -- transformation  together with $ \k$ spinor forms
$ \a(\k)$ -- superdiffeomorphism parameter, and the second $ \l$
plays the role of the superpartner of $ \th$ in respect to these
transformations.

The first papers on twistor like actions were \cite{vz}, \cite{stv},
\cite{stvz}. Afterwards a number of authors gave their contributions
\cite{sphet,ghs,bz,10D,gs2,11D,bpstv}.

Different but practically coinciding variants of twistor -- like action
for $ D=3,4,6,10$ superparticle and superstring were elaborated.
All of them solve the problem of the $\kappa$-- symmetry in the same way
as it has been explained above.

At the same time some basic problems
have not been solved satisfactory in the known versions of the
approach both from the
aesthetic and practical point of view. For instance, for constructing
the superfield action one should use superfield Lagrange multipliers.
Though some
of their components can be identified (on the mass shell) with the
momentum density and the tension of the super--p--brane, in general, the
geometrical and physical meaning of Lagrange multipliers is obscure.
Moreover, in a version suitable for the description of D=10, 11 objects
\cite{10D,11D}
 their presence in the action gives rise to some new symmetries
which turn out to be infinite reducible themselves, so that the problem
which we fighted in the conventional Green--Schwarz formulation
reappeared in a new form in the twistor--like formulation. Another point
concerning Lagrange multipliers is that in the superfield
formulation of D=10 type II superstrings \cite{gs2} and a D=11, N=1
supermembrane \cite{11D} Lagrange multipliers become propagative redundant
degrees of freedom which may spoil the theory at the quantum level.

Concluding this part of the lecture we note that the twistor-- like
approach has allowed to construct world -- sheet superspace formulation of
superparticle and superstring actions, which has together with nice
features some mentioned above drawbacks.

The latter are connected with introduction of Lagrange multipliers into
the theory, which in its turn is due to the fact that the corresponding
actions are defined by means of the Beresin integral, which
is not an exterior differential form and has no clear geometrical
meaning.

In the next part of the lecture we will see that the application of E.
Cartan's methods of differential geometry to the twistor approach makes
the theory more selfconsistent.

\section{The Generalized Action Principle}

Geometrical features of classical and quantum field theories as well as of
the theory of (super)particles and (super)strings can be the most
adequately expressed by using the exterior differential forms, firstly
proposed by E. Cartan.

E. Cartan's method has began to be widely used in the elementary particle
physics in the end of the 60's when general methods of phenomenological
Lagrangians and gauge fields were elaborated.

The advantage of applying exterior differential forms to construct action
for Goldstone fermions and supergravity was demonstrated as early as in
1972~--~1973 years
\cite{va}--
\cite{supergrrew}
\footnote{At that time E. Cartan's geometric ideas and his method of
the exterior differential forms were not well known to the particle
theorists. The most striking example is the work of
S. Ferrara, D. Freedman and P. Van Niewenhuizen \cite{ferrara}.
The authors, being  aquainted with our articles \cite{supergr},
nevertheless, used the cumbersome second order formalism to rediscover
$N=1$ supergravity action.}.

An important step in applying E. Cartan's method to supergravity was made
in the end of the 70's when Regge, Ne'eman, D'Auria and Fre had formulated
the Generalized Action Principle together with the rheonomy principle
\cite{rheo},
which allowed to extrapolate in some cases a component formalism of
supergravity on the superspace.

Here we apply GAP together with a new principle, which we name as
rheotropy principle (and which is an analog of rheonomy one) to formulate
the procedure of minimal embedding of super world sheet of superstrings and
supermembranes into the target superspace \cite{bpstv}.

We formulate the GAP for these cases as follows \cite{bsv}:
\begin{description}
\item[i)]
The action is a differential (p+1)--superform integrated over the
(p+1)--dimensional bosonic submanifold ${\cal
M}_{p+1}~:\{(\xi^m,\eta);~\eta=\eta(\xi)\}$ on the world supersurface
$$
S=\int\limits^{}_{{\cal M}^{p+1}}{\cal{L}}_{p+1}.
$$
The Lagrangian ${\cal{L}}_{p+1}$ is constructed  out of the vielbein
differential one--forms
in the target superspace and world supersurface ({\sl a priori} considered
as independent) by use of the exterior product
of the forms without any application of the Hodge operation.

To get the superfield equations of motion both the coefficients of
the forms and the world supersurface  are varied.
\item[ii)]
The intrinsic geometry of the world supersurface is not {\sl a priori}
restricted by any superfield constraints.
All the constraints and the geometrodynamical
conditions are obtained as variations of the action.
\item[iii)]
The field variations of the action gives two kinds of relations:\\
{}~~~1) relations between target superspace and world
supersurface vielbeins which orientate them along one another and
are the standard relations of surface
embedding theory; we call them ``rheotropic'' conditions
(`rheos' is `current' and `tropos' is `direction, rotation' in
Greek) \\
{}~~~2) dynamical equations causing the embedding to be minimal.\\
Only the latter equations put the theory on the mass shell.
\item[iv)]
The theory is superdiffeomorphism invariant off the mass shell if for
the action to be independent on the surface ${\cal M}_{p+1}$ (i.e.
$d{\cal{L}}_{p+1}=0$) only the rheotropic relations are required, and the
latter do not lead to equations of motion.

\end{description}

With all these points in mind we propose a super--p--brane action in the
following form:
$$ S_{ D,p} =-{{(-1)^p}\over{p!}}
\int\limits^{}_{{\cal M}_{p+1}}
(E^{a_0}
e^{a_1}...e^{a_{p}}
 - {{p}\over{(p+1)}}
e^{a_0} e^{a_1}... e^{a_{p}})\varepsilon_{a_0a_1...a_{p}}$$
$$ \pm(-i)^{{{p(p-1)}\over 2}-1}{p\over{(p+1)!}}
\int\limits^{}_{{\cal M}_{p+1}}
\sum^{p+1}_{k=0}
\Pi^{\underline{m}_p} \ldots
\Pi^{\underline{m}_{k+1}} dX^{\underline{m}_k}
\ldots dX^{\underline{m}_1}
d\Theta \Gamma_{\underline{m}_1\ldots
\underline{m}_p} \Theta
$$
where the exterior product of the differential forms is implied,
$ \varepsilon_{a_0a_1...a_{p}}$ is the unit antisymmetric tensor on
$ {\cal M}_{p+1}$.
$ e^a(\xi,\eta)$ are the bosonic vector components of a world
supersurface vielbein one--form $ e^A=(e^a,e^{\a p})$.  The external
differential $ d$
$$ d=e^aD_a+e^{\a p}D_{\a p}$$
with $ D_a,~D_{\a p}$ being world--supersurface covariant derivatives.
$$
\Pi^{\underline{m}}=dX^{\underline{m}}-id\Th\G^{\underline{m}}\Th,\qquad
d\Th^{\underline{\mu}}$$
is a pullback onto world supersurface of the vielbein forms
in flat target superspace.

$$ E^{\underline a}=
\Pi^{\underline m}u_{\underline m}^{\underline a},
$$

$ u^{\underline{a}}_{\underline{m}}$ are the fibre coordinates of the
target space vector vielbein fibre bundle, usually called in physical
literature as vector Lorentz harmonics \cite{sok}; $u^{a}_{\underline{m}}$
$(a=0,\dots,p)$ and $u^{i}_{\underline{m}}$ $(i=p+1,\dots,D)$ are its
subsets.

The vector Lorentz harmonics can be expressed by means of the spinor
Lorentz harmonics \cite{ghs,bz} by the relations
\begin{eqnarray}\label{uvv}
 \d_{qp} (\gamma_a )_{\a \b}
u^{~a}_{\underline{ m}} =
v_{\a q}
\Gamma_{\underline{ m}}
v_{\b p} ,
\qquad
\d_{\dot{q} \dot{p}} (\gamma_{a})^{\a \b}
u^{a}_{\underline{ m}} =
v^{\a }_{\dot q}
\Gamma_{\underline{m}}
v^{\b }_{\dot p}
\qquad  \nn
 \d^{\a}_{\b} \gamma ^{i}_{q \dot p}
u^{i}_{\underline{ m}}
= v_{\a q}
\Gamma_{\underline{ m}}
v^{\b }_{\dot p}
\end{eqnarray}

\section{Generalized action principle
for heterotic superstring}

The action for $ D=3,4,6,10$ heterotic string has the form
$$ {\cal S} = \int_{{\cal M}_{2}} ~~{\cal L}$$
\begin{eqnarray}\label{13}
{\cal L}
 = -{1 \over 2 \sqrt{\a^\prime}}
\left(
E^{++} e^{--} - E^{--} e^{++} + e^{--} e^{++} \right)
+ {\cal L}_{WZ}
\nn
{\cal L}_{WZ}
 = -{i \over \sqrt{\a^\prime}}
\Pi^{\underline{m}}
d\Th\G_{\underline{m}} \Th,
\end{eqnarray}
where $\a^\prime$ is the Regge slope parameter (inverse string tension).

In \p{13}
$
e^a(\xi,\eta) \equiv (e^{++}(\xi,\eta), e^{--}(\xi,\eta) )
$
are bosonic vector zweinbein
one--forms. The complete basis (supervielbein) contains also $(D-2)$ fermionic
$1$--forms $e^{+q}(\xi , \eta)$ :
\begin{equation}\label{c}
e^A=(e^{++}, e^{--}, e^{+q}),
\end{equation}
the latter are not involved into the construction of the action.
The external differential $d$ is:
\begin{equation}\label{d}
d=e^{\pm\pm}{\cal{D}}_{\pm\pm} + e^{+ p}{\cal{D}}_{+ p}
\end{equation}
with ${\cal{D}}_{\pm\pm},~{\cal{D}}_{+ q}$ being
world supersurface covariant derivatives.

\begin{equation}\label{pif}
\Pi^{\underline{m}}=dX^{\underline{m}}-id\Th\G^{\underline{m}}\Th,\qquad
d\Th^{\underline{\mu}}
\end{equation}
and
$$E^{\underline A} \equiv ( E^{++}, E^{--}, E^{\underline \a})$$
\begin{eqnarray}\label{rep_ind}
E^{\pm\pm} \equiv
\Pi^{\underline{m}}u_{\underline{m}}^{\pm\pm} , \qquad
E^{i} \equiv
\Pi^{\underline{m}}u_{\underline{m}}^{~i} , \qquad
E^{\underline \a} \equiv
d\Theta^{\underline{\mu}} v_{\underline{\mu}}^{~\underline \a} , \qquad
\end{eqnarray}
Moving frame vectors
$u_{\underline{m}}^{\pm\pm}(\xi,\eta),
{}~u_{\underline{m}}^{i}(\xi,\eta)$
are naturally composed of the spinor components (Lorentz harmonics)
\cite{bz}
$v_{\underline{\mu}}^{~\underline{\a}} =
( v^{~+}_{\underline{\mu} q}~,~v^{~-}_{\underline{\mu}\dot q} )
$
We present the corresponding expressions for the simplest $D=3$ case
\begin{eqnarray}\label{3D}
u^{++}_{\underline{ m}}
\Gamma^{\underline{ m}}_{\underline{ \mu} \underline{ \nu}}
= 2 v^{~+}_{\underline{\mu} } v^{~+}_{\underline{\nu} } , \qquad
u^{--}_{\underline{ m}}
\Gamma^{\underline{ m}}_{\underline{ \mu} \underline{ \nu}}
= 2 v^{~-}_{\underline{\mu} }
v^{~-}_{\underline{\nu} } , \qquad
u^{~\perp}_{\underline{ m}}
\Gamma^{\underline{ m}}_{\underline{\mu} \underline{\nu}}
= v^{~+}_{\underline{\mu} } v^{~-}_{\underline{\nu} }
+ v^{~-}_{\underline{\mu} } v^{~+}_{\underline{\nu} }
, \qquad    \nn
v^{-\underline{\mu}} v^{+}_{\underline{\mu}} \equiv
\e^{\underline{ \mu} \underline{ \nu}}
v^{-}_{\underline{\nu}}
v^{+}_{\underline{\mu}}
\equiv
v^{-\underline{\mu}}
\e_{\underline{ \mu} \underline{ \nu}}
v^{+}_{\underline{\nu}} = 1.
\end{eqnarray}

The last term in \p{13} is Wess-Zumino term,
its coefficient being fixed by
the requirement that when the action \p{13} is restricted to the
component formulation of the superstring
the resulting action has local
$\kappa$--symmetry.

To get the equations of motion, constraints and rheotropic relations let us
vary the action \p{13}
and then project the result on the world -- sheet supervielbein \p{c}

\begin{equation}\label{e}
{\d S\over{\d e^{\pm\pm}}} = 0 ~~~\Rightarrow~~~
E^{\mp\mp} \equiv \Pi^{\underline{m}}
u_{\underline{m}}^{\mp\mp} = e^{\mp\mp }
\end{equation}

\begin{equation}\label{u0}
u^{i}_{\underline{m}} {\d S\over{\d u^{\pm\pm}_{\underline{m}} }} = 0
{}~\Rightarrow ~ E^{i} e^{\mp\mp} = 0
\end{equation}
which means
\begin{equation}\label{u}
E^{i} \equiv
\Pi^{\underline{m}}
u_{\underline{m}}^{i} = 0
\end{equation}
on the world supersurface.
Eqs. \p{e} and \p{u} are part of the
rheotropic relations which define tangent and outer directions of the
vector target space vielbein.
They can be rewritten in the following form
\begin{equation}\label{pi}
\Pi^{\underline{m}} = {1 \over 2}
(e^{++} u^{--\underline{m}} + e^{--}  u^{++\underline{m}} )
\end{equation}

Eq. \p{pi} means that the embedding of the world supersurface satisfies the
geometrodynamical condition
\begin{equation}\label{geomd}
\Pi^{\underline{m}}_{+q}={\cal{D}}_{+q}X^{\underline{m}}
-i{\cal{D}}_{+q}\Th\G^{\underline{m}}\Th=0
\end{equation}
as well as the equations
\begin{equation}\label{virasoro}
{\Pi_{\pm\pm}}^{\underline{m}}\equiv {\cal{D}}_{\pm\pm}
X^{\underline{m}} - i {\cal{D}}_{++} \Th\G^{\underline{m}} \Th =
u_{\pm\pm}^{\underline{m}}
\end{equation}
which are  "the square roots" of the Virasoro
constraints
\footnote{Due to the presented below identification of twistor
${\cal D}_{+q} \Th^{\underline{m}}$ and Lorentz harmonic
$v^{-\underline{m}}_q$ superfields, one of this relations can be
identified with the so--called  "twistor constraint"
\cite{stv,sphet,gs2,10D,11D}}
$({\Pi_{++}}^{\underline{m}})^2 = 0, ~
({\Pi_{--}}^{\underline{m}})^2 = 0$.

The other equations of motion are got by varying $\d{X^{\underline{m}}}$
and $\d\Th^{\underline{\mu}}$
\begin{equation}\label{x}
{\d S\over{\d X^{\underline m}}}
= 0
{}~\Rightarrow~
d( u^{++}_{\underline{m}} e^{--} - u^{--}_{\underline{m}} e^{++})
- 2i d\Th \G_{\underline {m}} d\Th = 0,
\end{equation}
\begin{equation}\label{th}
{\d S\over{\d\Th^{\underline{\mu}}}}=0
{}~\Rightarrow~
d\Th^{\underline{\mu}}\G^{\underline{m}}_{\underline{\mu\nu}}
(u^{--}_{\underline{m}} e^{++} - u^{++}_{\underline{m}} e^{--}
+ 2 \Pi_{\underline{m}}) = 0
\end{equation}
Eq.\p{x} can be reduced to straightforward superfield
generalization of the Green--Schwarz equation when \p{virasoro} is taken
into account. It can be proved that this equations are satisfied
identically due to 'rheotropic relations' and Eq.\p{th}.
Eq.\p{th} contains the following equation for $\Th$
\begin{equation}\label{theqm}
{\cal{D}}_{--}\Th^{\underline{m}}v^{-}_{\underline{m}\dot{q}}=0
\end{equation}
together with the rheotropic relation
\begin{equation}\label{thrheo}
{\cal{D}}_{+q}\Th^{\underline{m}}v^{-}_{\underline{m}\dot{q}}=0
\end{equation}
for the spinor target space vielbein.
\medskip

As the Lagrangian of the generalized action coincides with that of the
component formalism the leading terms in the superfield expansion in
$\eta$'s coincide with those of component formalism.

Equations with ${\cal{D}}_{+q}$ allow to spead the solution to the whole
superworld surface.

\medskip

Now we make the last essential remark.

As it can be easily verified the requirement {\bf iv)} in the definition of
GAP is satisfyed if the rheotropic relations \p{e}, \p{u} and
\p{thrheo} are taken into account.
If one changes the ratio of the
coefficients before the first and second terms in \p{13} the requirement
{\bf iv)} will not be satisfied. This fact expresses the connection of the
$\kappa$ -- symmetry with superdiffeomorphism symmetry
on the super world -- sheet.

\section{Resum\'e.}

The twistor -- like superfield approach to superstrings and supermembranes
is revised, with emphasis upon its drawbacks. It is shown that the
newly proposed generalized action principle which is akin to that used in
the group manifold approach to supergravity \cite{rheo} reproduces all the
"good results" of the twistor approach and, because there are no Lagrange
multipliers in the theory, it is free of its difficulties which are due to
their presence in previous formulations.

\vspace{0.8cm}
{\large\bf Acknowledgements.}

I wish to thank the organizers of the "SUSY -- 95" Workshop for
participation in very stimulating meeting. This work was supported in part
by ISF Grant RY 9200 and by the INTAS Grants 93 -- 127, 93 -- 493, 93 --
633.


{\small
\begin{thebibliography}{99}

\bibitem{rns}
P. Ramond \PRD 3 1971 2415 \\
A. Neveu and J.H.  Schwarz \NPB 31 1971 86 ; \PRD 4 1971 1109

\bibitem{gs}
M. Green and J. Schwarz, \NPB 243 1984 285.

\bibitem{veccia}
L. Brink, P. Di Vecchia and P. Howe, \PLB 65 1971 471

\bibitem{bs}
L. Brink and J. Schwarz, \PLB 101 1981 310.

\bibitem{stv}
D. Sorokin, V. Tkach and D. V. Volkov, {\sl Mod. Phys. Lett.} {\bf A4}
(1989) 901.
\bibitem{k}
J. De Azcarraga and J.Lukiersky, {\sl Phys. Lett.} {\bf B113} (1982)
170.\\
W.Siegel, {\sl Phys. Lett.} {\bf B128} (1983) 397.

\bibitem{gs92}
A. Galperin and E. Sokatchev, {\sl Phys. Rev.} {\bf D46} (1992) 714.

\bibitem{vz}
D. V. Volkov and A. Zheltukhin, \JETPL 48 1988 61;
\LMP 17 1989 141; \NPB 335 1990 723.

\bibitem{stvz}
D. Sorokin, V. Tkach, D. V. Volkov and A. Zheltukhin, {\sl Phys. Lett.}
{\bf B216} (1989) 302.

\bibitem{sphet}
N. Berkovits, {\sl Phys. Lett.} {\bf
232B} (1989) 184; {\bf 241B} (1990) 497;
\\ E. Ivanov and A. Kapustnikov {\sl Phys. Lett.}, {\bf B267} (1991) 175.
\\ M. Tonin, {\sl Phys. Lett.} {\bf B266} (1991) 312; \\
F. Delduc and E. Sokatchev, {\sl Class. Quantum Grav.} {\bf 9} (1992)
361.\\
S. Aoyama, P. Pasti and M. Tonin, {\sl Phys. Lett.} {\bf B283}
(1992) 213;
\\ M. Tonin, {\sl Int. J. Mod.Phys} {\bf 7} (1992) 613;
\\ F. Delduc, E. Ivanov and E. Sokatchev, \NPB 384 1992 334;
\\ D Volkov, {\sl Spinors, Twistors, Clifford Algebras and Quantum
Deformations, Z.Oziewicz et all. (eds.), Kluwer Academic Publishers} 1993
109.


\bibitem{ghs}
A. Galperin, P. Howe and K. Stelle, \NPB 368 1992 248;
A. Galperin, F. Delduc and E. Sokatchev, \NPB 368 1992 143.

\bibitem{bz}
I. A. Bandos, {\sl Sov. J. Nucl. Phys.} {\bf 51} (1990) 906;\\
I. A. Bandos and A. A. Zheltukhin ,
{\sl Phys. Lett.} {\bf B261} (1991) 245;
 {\sl Theor. Math. Phys.} {\bf 88} (1991) 358;
 {\sl Fortschr. Phys.} {\bf 41} (1993) 619--676. \\
 {\sl Phys. Lett.} {\bf B288} (1992) 77;
 {\sl Phys. Atom. Nucl.} {\bf 56} (1993)
 113--121;
{\sl Phys. Part. Nucl.} {\bf 25} (1994) 453--477. \\
 {\sl Int. J. Mod. Phys.} {\bf A8} (1993) 1081-- 1091.
{\sl Class. Quantum Grav.} {\bf 12} (1995) 609--626
 {\bf (hep-th/9405113)}.

\bibitem{gs2}
A. Galperin and E. Sokatchev, \PRD 48 1993 4810.

\bibitem{10D}
F. Delduc, A. Galperin, P. Howe and E. Sokatchev, {\sl Phys. Rev.} {\bf
D47} (1992) 587.

\bibitem{11D}
 P. Pasti and M. Tonin, \NPB 418 1994 337.\\ E. Bergshoeff and E.
Sezgin, \NPB 422 1994 329.\\ E. Sezgin, Preprint CTP TAMU--58/94, November
1994; {\bf (hep--th 9411055)}.


\bibitem{bpstv}
I. Bandos, P. Pasti, D. Sorokin, M. Tonin and D. Volkov,
\NPB 446 1995 79-- 118.
{\bf (hep-th/9501113)}.

\bibitem{va}
D.V. Volkov and V.P. Akulov, \JETPL 16 1972 438;
\PLB 46 1973 109; \TMP 18 1974 39. \\

\bibitem{supergr}
D.V. Volkov and V.A. Soroka, \JETPL 18 1973 312;
\TMP 20 1974 291.

\bibitem{supergrrew}
D. Volkov, {\sl Supergravity before 1976},
{\bf hep-th/9410024}, to be published in: {\sl ''Proc. of Int. Conf. on the
History of Original Ideas and Basic Discoveries in Particle Physics.
Erice, Italy, 1994}

\bibitem{ferrara}
S. Ferrara, D. Freedman and P. van Niewenhuizen \PRD 13 1976 3214.

\bibitem{rheo}
Y. Nieman and T. Regge, \PLB 74 1978 31, {\sl Revista del Nuovo Cim.}
1 1978 1;\\
R. \'D Auria, P. Fr\'e and T. Regge, {\sl Revista del Nuovo Cim.}
3 1980 1;\\
L. Castellani, R. \'D Auria, P. Fr\'e. ``Supergravity and superstrings, a
geometric perspective'', World Scientific, Singapore, 1991 (and
references therein).

\bibitem{bsv}
I. Bandos, D. Sorokin and D. Volkov,
       {\sl Phys.Lett.}{\bf B352} (1995) 269
       {\bf (hep-th/9502141)}.

\bibitem{sok}
E. Sokatchev , {\sl Phys. Lett.} {\bf B169} (1987) 209;
\CQG 4 1987 237.

\end{thebibliography}   }
\end{document}